\documentclass[a4paper,11pt]{article}
\usepackage{jcappub}
\usepackage{amsmath,amssymb,amsthm}

\usepackage{mathrsfs}     % \mathscr
\usepackage{empheq}       % boxed equations
\usepackage{mathtools}    % dcases, etc.
\usepackage{bm}           % \bm bold math
\usepackage{dcolumn}      % decimal-aligned table columns
\usepackage{cleveref}     % \Cref / \cref (load AFTER hyperref)
\usepackage{soul}         % \hl highlighting
\usepackage{comment}      % comment environment
\usepackage{booktabs}
\usepackage{fontawesome5} % \faGithub icon
\usepackage{orcidlink}    % \orcidlink{XXXX-...}

\definecolor{darkblue}{rgb}{0.0,0.0,0.5}
\hypersetup{
    linkcolor=darkblue,
    citecolor=blue,
    urlcolor=darkblue,
}

\newcommand{\be}{\begin{equation}}
\newcommand{\ee}{\end{equation}}
\newcommand{\bs}{\begin{split}}
\newcommand{\bea}{\begin{eqnarray}}
\newcommand{\eea}{\end{eqnarray}}
\newcommand{\al}{\alpha}
\newcommand{\alb}{\alpha_B}
\newcommand{\alk}{\alpha_K}
\newcommand{\dl}{\delta}

\newcommand{\ode}{\Omega_{\rm de}}

\newcommand{\wde}{w_{\rm de}}
\newcommand{\wtot}{w_{\rm tot}}

\newcommand{\rde}{\rho_{\rm de}}
\newcommand{\pde}{P_{\rm de}}

\newcommand{\mpl}{M_{\rm Pl}^2}

\newcommand{\fe}{f_{\rm early}}
\newcommand{\we}{w_{\rm early}} 
\newcommand{\nearly}{n_{\rm early}}
\newcommand{\nlate}{n_{\rm late}}

\newcommand{\github}[1]{\href{#1}{\faGithub}}

\title{Fresh Look at Shift Symmetric Modified Gravity}

\author[a]{Rodrigo Calder\'on\orcidlink{0000-0002-8215-7292}}
\author[b]{Eric V.\ Linder\orcidlink{0000-0001-5536-9241}}

\affiliation[a]{CEICO, Institute of Physics of the Czech Academy of Sciences,
  Prague, Czech Republic}
\affiliation[b]{Berkeley Center for Cosmological Physics \& Berkeley Lab,
  University of California, Berkeley, CA 94720, USA}

\emailAdd{calderon@fzu.cz}
\emailAdd{evlinder@lbl.gov}

\abstract{Shift symmetric Horndeski gravity can be a 
simple, well posed theory with protection against quantum 
corrections. Overturning conventional thought that such a 
theory cannot 
cross the phantom divide ($w=-1$) we present both analytic 
and numerical proof that it can. 
We carry out a detailed dynamical systems analysis showing 
the evolution of ghost free and stable regions within the 
model family, identify the fixed point and dynamical 
attractor structure, and derive the interaction of the 
$w=-1$ curve with these phase space structures. 
This reveals two mechanisms for cosmologies crossing the phantom 
divide while preserving shift symmetry.} 

\begin{document}

\maketitle

%%%%%%%%%%%%%%%%%%%%%%%%%%%%% 
\section{Introduction}

Dark energy from a non-interacting scalar field, whether 
with a canonical kinetic structure like quintessence or 
a noncanonical one like k-essence, must evolve with an 
equation of state ratio $w$ wholly on one side or the other 
of the cosmological constant value $w=-1$. This boundedness is 
referred to as the phantom divide. One can cross $w=-1$ by 
modifying gravity away from general relativity, through a 
scalar-tensor theory. 

Horndeski gravity is a general 
scalar-tensor allowing noncanonical kinetic structure, 
nonminimal coupling to matter, and braiding between the scalar 
and tensor aspects. The terms associated with these effects 
can be both field and field derivative dependent. Nonminimal 
coupling -- often problematic to fit observations 
\cite{2506.02122,2407.02558,DESI-MG:2024,2409.17019,2605.12415,Desmond:2026lds} 
-- relies on 
field dependence. Moreover, field dependent terms run into 
issues with high energy (early universe) quantum loop 
corrections \cite{Kolda:1998wq,Peccei:2000rz,Doran:2002bc}. Therefore there is great interest in terms that 
rely only on field derivatives, and hence have protective 
shift symmetry. However, common wisdom in the literature is 
that an unbroken shift symmetric Horndeski theory that is both 
ghost free and perturbatively stable cannot cross 
the phantom divide \cite{tsuji,2509.17586,2103.11195,1008.0048,0407107}, as observations appear to prefer \cite{DESI-Adame2024,DESI-calderon2024,DESI-lodha2025,DES:2026jmi}. 

We revisit this issue and show that quite simple shift symmetric 
theories can in fact cross the phantom divide (though we do not 
here attempt to fit the data in detail). Taking a dynamical 
systems approach detailed in Section~\ref{sec:dyn}, we identify 
the regions of phase space free of ghosts and instabilities in 
Section~\ref{sec:phase}, together with the fixed points and 
attractor structure. Overlaying the $w=-1$ condition reveals the 
analytic conditions, discussed in Section~\ref{sec:w1}, needed for crossing the phantom divide. Section~\ref{sec:cross} 
presents two mechanisms for crossing and presents the full 
numerical evolution solutions. Section~\ref{sec:concl} concludes.

%%%%%%%%%%%%%%%%%%%%%%% 
\section{Dynamical System for Shift Symmetry} \label{sec:dyn} 

The Lagrangian for shift symmetric Horndeski gravity with a 
luminal speed of tensor perturbation propagation is 

\be 
\mathcal{L}=\frac{1}{2}R+K(X)-G_3(X)\,\Box\phi \ , 
\ee 
where $X=(1/2)\dot\phi^2$ and we set $\mpl=1$. Note that there is no potential: we 
do not allow a cosmological constant because, while shift 
symmetric, it suffers similar quantum correction issues. 

The Friedmann and scalar field evolution equations are 
\bea 
3H^2&=&\rho_m+\rde\ ,\label{eq:h2}\\  
-2\dot H&=&\rho_m+P_m+\rde+\pde\ ,\label{eq:hdot}\\  
0&=&\ddot\phi\,\left[K_X+2XK_{XX}+6H\dot\phi g_X\right]
+3H\dot\phi K_X+6g\left(\dot H+3H^2\right)\ , \label{eq:fd} 
\eea 
where a subscript $X$ denotes a derivative with respect to $X$, 
we write $g\equiv XG_{3X}$, and we identify the effective dark 
energy density and pressure 
as 
\bea 
\rde&=&-K+2XK_X+6H\dot\phi g\ ,\label{eq:rho}\\  
\pde&=&K-2g\ddot\phi\ . \label{eq:pde} 
\eea 

We define a phase space using the variables representing  
kinetic and braiding contributions to the dark energy density, 
\bea 
x_k &=& \frac{K}{3H^2\mpl}\\ 
x_g &=& \frac{6H\dot\phi\,g}{3H^2\mpl}\ , 
\eea 
here writing $\mpl$ explicitly to show these are dimensionless quantities, and  
\bea 
\ode&=&(2n-1)\,x_k+x_g\\ 
w\ode&=&x_k-x_gx_\phi\ , 
\eea 
and $n=XK_X/K$, while 
\bea 
x_\phi &\equiv& \frac{\ddot\phi}{3H\dot\phi}=\frac{X'}{6X} \quad;\quad \al x_\phi=
-6nx_k-\frac{3}{2}x_g(1-x_k) \label{eq:alxphi}\\ 
x_H &\equiv& -3(1+\wtot)=-3\left(1+x_k-x_g x_\phi+\frac{\Omega_r}{3}\right)\\ 
\al &\equiv& \alk+\frac{3}{2}\alb^2\\ 
\alk &=& \left[6n(2n-1)+12Xn_X\right]x_k+6mx_g\\ 
\alb &=& x_g\ . 
\eea 
A prime denotes $d/d\ln a$, while 
$m\equiv Xg_X/g$ and we will generally take $n_X=0$ 
(i.e.\ a monomial $K\sim X^n$) and $\Omega_r=0$ (focus on the 
matter dominated era and later) unless otherwise stated. The 
$\al_i$ quantities are often referred to as property functions, 
with $\alk$ the kineticity and $\alb$ the braiding parameter \cite{bellsaw}. 

The equations of motions -- Eqs.~(\ref{eq:h2}), (\ref{eq:hdot}), 
(\ref{eq:fd}) -- become a dynamical coupled system of autonomous first order 
differential equations in a $x_k$--$x_g$ phase space: 
\bea 
x_k' &=& x_k\left[\,6n\,x_\phi - x_H \right] \\
x_g' &=& x_g\left[\,3(1+2m)\,x_\phi - \frac{x_H}{2} \right] \quad \to\quad \ x_g\left[\frac{3(2n+1)\,x_\phi}{2} - \frac{x_H}{2} \right]\ . 
\eea 
In the presence of shift symmetry, a conserved charge quantity 
appears and there is a fixed relation $m=(2n-1)/4$ as discussed 
in \cite{Calderon:2026hbr}, and this is indicated by the right arrow. 
There is a subtlety if 
$n$ and $m$ are not constants, but unless otherwise stated 
(see Sec.~\ref{sec:cross} below) we 
take them to be constant, i.e.\ $K\sim X^n$ and $G_3\sim X^m$. 
This gives a particularly simple theory with few free parameters, 
and will be our main focus.

%%%%%%%%%%%%%%%%%%%%%%% 
\section{Phase Space Analysis} \label{sec:phase} 

The first thing to note is that the lines $x_k=0$ and $x_g=0$ 
are invariant; the quantities $x_k$, $x_g$ cannot flip sign during evolution 
and so the trajectories must stay within the initial quadrant 
of the phase space. 

To check the health of the theory, we require the no ghost 
and stability conditions, respectively 
\bea 
\al&=&\alk+\frac{3\alb^2}{2}=6n(2n-1)\,x_k+\frac{3(2n-1)}{2}\,x_g+\frac{3}{2}\,x_g^2 \ge 0\\ 
\al c_s^2&=&\frac{1}{2}(2-x_g)(x_g-x_H)+x_g'-3(1-\ode)\ge 0\ . 
\eea 

Dynamical system analysis provides the fixed points and 
eigenvalues of the coupled evolution equations. 
Table~\ref{tab:fp} presents the results. We immediately 
see that $n<0$ will give a ghost. Since the dark energy 
equation of state at early times is $w=1/(2n-1)$ due to charge conservation 
\cite{2512.03139,Calderon:2026hbr}, this means that initial conditions with 
$w\in(0,1)$ cannot give a viable theory with late time dark 
energy domination within this model. For $0<n<1/2$ -- a phantom early dark energy 
equation of state -- we can have a healthy theory, with for example 
a de Sitter attractor. The case of $n>1/2$ is problematic since 
then the early dark energy equation of state is $w>0$ and this would dominate over 
matter at early times (and over radiation for $1/2<n<2$). 
Furthermore, $n>1$ has no stable attractor. 
Therefore we focus on $0<n<1/2$, with phantom behavior at early 
times, as in \cite{Calderon:2026hbr}.

%%%%%%%%%%%%%%%%%%%% 
\begin{table}[tb]
\centering
\small
\begin{tabular}{llccccccl}
\toprule
Notation & $(x_k,x_g)$ & $\ode$ & $w$ & $f$ & $\alpha$ & $c_s^2$ & eigenvalues & type\\
\midrule
M & $(0,0)$ & $0$ & --- & --- & $0$ & --- & $\dfrac{-3}{2n-1}$ (twice) & repeller\\[1.1ex]
K & $\left(\dfrac{1}{2n-1},0\right)$ & $1$ & $\dfrac{1}{2n-1}$ & $0$ & $6n$ & $\dfrac{1}{2n-1}$ & $\dfrac{3}{2n-1},\ \dfrac{-3}{2(2n-1)}$ & saddle\\[1.1ex]
B & $(0,1)$ & $1$ & $\dfrac{1}{2n}$ & $1$ & $3n$ & $\dfrac{8n+3}{12n^2}$ & $\dfrac{3}{2n}$ (twice) & repeller\\[1.1ex]
dS & $(-1,2n)$ & $1$ & $-1$ & $2n$ & $3n$ & $\dfrac{2(1-n)}{3}$ & $-3$ (twice) & attractor\\
\bottomrule
\end{tabular}
\caption{Fixed points of the dynamical system, with their observables, eigenvalues, and classification. 
$M$ is the early matter dominated point, $K$ has dark energy 
dominated by its kinetic term, $B$ by its braiding term, and 
$dS$ is a de Sitter final state. 
} 
\label{tab:fp}
\end{table}

\begin{figure}
    \centering
    \includegraphics[width=\linewidth]{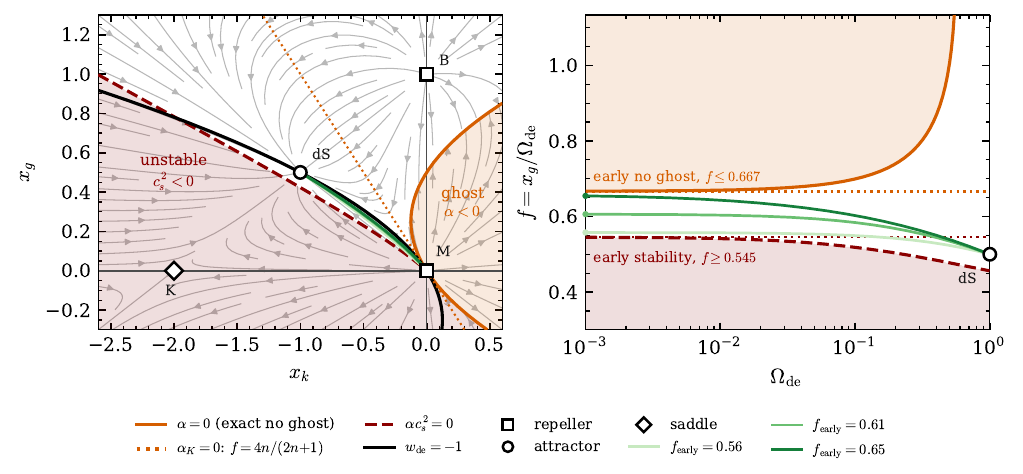}\\
    \caption{Exact 2D flow in the $(x_k,x_g)$-plane for 
    $n=1/4$ and $m=(2n-1)/4$ (the case considered in \cite{Calderon:2026hbr}). Trajectories can safely reach the asymptotic dS attractor without running into the ghost or instability regions. 
The left panel shows the $x_k$--$x_g$ dynamical space with 
streamlines, while the right panel maps this onto the $f$--$\ode$ 
space, where $f$ is the fraction of dark energy density due to 
the braiding term. Trajectories for three different values of 
$\fe$ are shown (green curves; in 
the left panel they lie almost on top of each other). The de Sitter attractor has $(x_k,x_g)=(-1,2n)$, or 
equivalently $(f,\ode)=(2n,1)$. 
    }
    \label{fig:flow-paperI}
\end{figure}

\begin{figure}
    \centering
    \includegraphics[width=\textwidth]{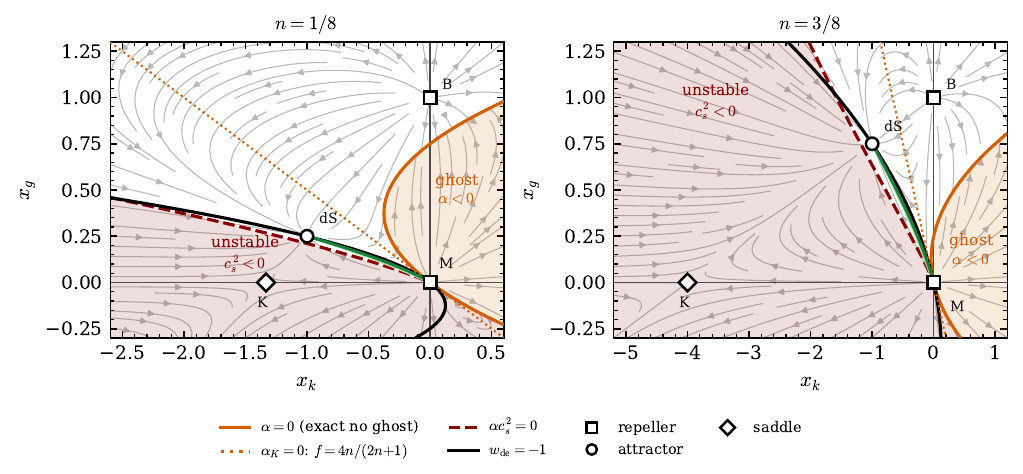}
    \caption{Exact 2D flow in the $(x_k,x_g)$-plane for 
    $n=1/8$ (left) and $n=3/8$ (right). Trajectories can safely reach the asymptotic dS attractor without running into the ghost or instability regions. 
    }
    \label{fig:flown125375}
\end{figure}

Figure~\ref{fig:flow-paperI} and Figure~\ref{fig:flown125375} show 
the phase space trajectories for $n=1/8$, $n=1/4$, $n=3/8$. 
Note how the ghost and instability regions shift across the 
phase space, but a clear healthy path always lies between 
the initial matter dominated state and the final de Sitter state. 
Also, since $x_g\equiv f\ode$ and $x_k=\ode(1-f)/(2n-1)$, the map $(x_k,x_g)\leftrightarrow(\ode,f)$ is a bijection, and
lines through the origin are curves of constant $f$. One can 
show directly from $x_g'$ that $f'\sim\mathcal{O}(\ode)$, i.e.\ 
that $f$ becomes constant at early times. Conditions on this 
early time $f$ limit for a healthy theory were derived in \cite{Calderon:2026hbr}. The right panel of 
Figure~\ref{fig:flow-paperI} illustrates that view of the 
phase space.

%%%%%%%%%%%%%%%%%%%%%%%% 

%%%%%%%%%%%%%%%%%%%%%%%% 

%%%%%%%%%%%%%%%%%%%%%%% 
\section{Analytic Conditions for Crossing the Phantom Divide} 
\label{sec:w1} 

While we have established a class of healthy theories, and the 
end states of their evolution, we have not yet addressed the 
crossing of the phantom divide. The trajectories can end up at 
$w=-1$ but can they cross $w=-1$ before reaching the de Sitter 
state? 

Consider the quantity $\al(\ode+w\ode)$. This will be zero 
whenever $w=-1$. Using Eq.~(\ref{eq:alxphi}) one finds 
\bea 
1+\wde&=&\frac{3\,Q}{2\,\alpha} \label{eq:wQ}\\ 
Q(x_k,x_g)&\equiv& x_g^2+2n\,x_g+8n^2x_k \ . \label{eq:Qxgxk}
\eea 
For $\ode\neq0$, $\wde=-1$ is therefore the phase space parabola $Q=0$, which passes through $M$ and
through de Sitter, $dS=(-1,2n)$. Wherever $\alpha>0$ then $\operatorname{sign}(1+\wde)=\operatorname{sign}Q$. 

Calling the value of $x_g$ where the trajectory hits $Q=0$ by $s$, we have there 
\bea 
x_k(Q=0)&=&\frac{-s(s+2n)}{8n^2}\\ 
\al(Q=0)&=&\frac{3s^2}{4n}\ . 
\eea 
Again this is only ghost free for $n>0$. 

The question ``does $\wde$ cross $-1$?'' is the question ``does the trajectory
cross the curve $Q=0$?''. The parabola splits the plane into two sides, $Q>0$ and $Q<0$,
which are exactly $\wde>-1$ and $\wde<-1$ wherever $\alpha>0$. To see whether, and in which
direction, a trajectory passes from one side to the other, follow the value of $Q$ along it.
Writing $Q(N)\equiv Q\big(x_k(N),x_g(N)\big)$, with time coordinate $N\equiv\ln a$, the chain rule gives
\begin{equation}
Q'=\frac{\partial Q}{\partial x_k}\,x_k'+\frac{\partial Q}{\partial x_g}\,x_g'
  =\nabla Q\cdot\big(x_k',x_g'\big)
  =8n^2\,x_k'+2\,(x_g+n)\,x_g'\ .
\end{equation}
Since $\nabla Q$ is normal to the curve $Q=0$, $Q'$ evaluated \emph{on} the curve measures the
component of the flow velocity that points across it. There are two possibilities at each
point of the curve:
\begin{itemize}
\item $Q'=0$: the velocity is tangent to the curve. If this held at \emph{every} point, a
      trajectory starting on the curve would stay on it. The curve would then be made of
      orbits (an invariant curve), and nothing could cross it.
\item $Q'\neq0$: the velocity has a component across the curve, so the trajectory through
      that point passes from one side to the other. If a trajectory is on the curve at time
      $N_1$, then $Q(N_1)=0$ and, to first order, $Q(N)\simeq Q'(N_1)\,(N-N_1)$. If
      $Q'(N_1)<0$, the trajectory has $Q>0$ just before $N_1$ and $Q<0$ just after: it
      crosses from $\wde>-1$ to $\wde<-1$. If $Q'(N_1)>0$, it crosses the other way.
\end{itemize}
To evaluate $Q'$ on the curve, substitute the flow equations into $Q'$ and use the
parametrisation $x_g=s$, $x_k=-s(s+2n)/(8n^2)$. On the curve $\alpha=3s^2/(4n)\neq0$ 
so the flow is regular there. The result is
\begin{equation}
Q'\big|_{Q=0}=-\frac{3\,(2n+1)}{8n}\;s\,(s-2n)^2\ . \label{eq:Qdot}
\end{equation}
It vanishes only at $s=0$ and $s=2n$, i.e.\ at $M$ and $dS$. That is expected: these are
fixed points, where the velocity itself is zero. At every other point of the curve $Q'\neq0$,
so the parabola is not an orbit, and every trajectory that reaches it crosses it there.
Moreover, the sign of \eqref{eq:Qdot} is the same at every point of a given branch, $s>0$ or
$s<0$. Each branch is therefore a one-way door: it can only be crossed in a single direction.

Finally, $x_g'$ is proportional to $x_g$, so the axis $x_g=0$ is itself an orbit
and no trajectory can cross it. A trajectory therefore keeps the sign of $x_g$ and can only
ever meet one branch of the parabola.

%%%%%%%%%%%%%%%%%%%%%%% 
\section{Crossing Mechanisms} \label{sec:cross} 

For a healthy theory giving eventual dark energy domination, 
Table~\ref{tab:fp} requires $0<n\le 1$. First consider the 
case $0<n<1/2$. At early times $1+w=2n/(2n-1)<0$, i.e.\ 
phantom, and so $Q<0$. Moreover for early time stability we 
require $x_g>0$ (see \cite{Calderon:2026hbr}). 
Now suppose the trajectory manages to 
reach the $Q=0$ curve, i.e.\ $w=-1$, before reaching the de 
Sitter asymptote. At this point $Q'\le0$ by Eq.~(\ref{eq:Qdot}). 
Thus it cannot cross $Q=0$ but either bounces off, or sticks 
to $w=-1$ if it is at the de Sitter fixed point $s=2n$. 

For the case $1/2<n\le 1$ (leaving aside for the moment issues of 
early dark energy domination), initially $w>0$ and $Q>0$. 
We still have $Q'<0$, so one can in theory cross from $w>-1$ to 
$w<-1$. At early times $x_\phi=-1/(2n-1)$ (as in \cite{Calderon:2026hbr}) 
and so $x_g'=-3x_g/(2n-1)$. Thus for $n>1/2$ we have $x_g'<0$ 
and it will not evolve away from a matter dominated ($\ode\ll1$, 
$x_g\ll1$) state. However, a loophole is to start from a 
state with large enough $x_g$, specifically near the braiding 
fixed point with early dark energy domination. 

\begin{figure}
    \centering
    \includegraphics[width=\linewidth]{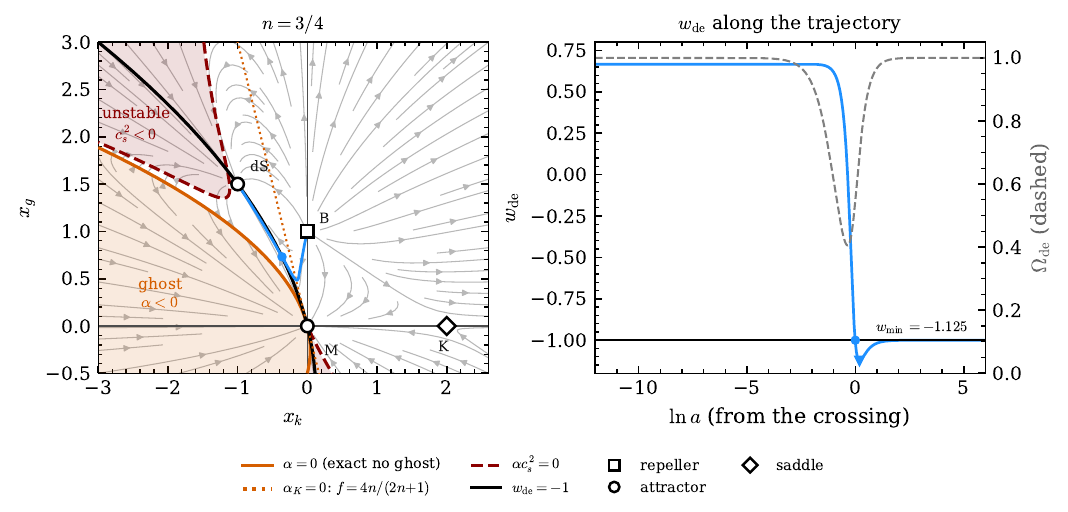}     
    \caption{
    A healthy phantom crossing for a shift symmetric model with $n=3/4$. [Left panel] The trajectory (blue) leaves
      the braiding point $B$, crosses the parabola $\wde=-1$ (black) once at the dot, and ends at
      $dS$ from the phantom side. [Right panel] $\wde$ (solid) and $\ode$ (dashed) along the trajectory,
      against $\ln a$ counted from the crossing. The trajectory starts at $\ode=1$, so it has no matter dominated era.}
      \label{fig:crossing_example}
\end{figure}

Figure~\ref{fig:crossing_example} plots the trajectory for $n=3/4$, 
and shows that indeed $w$ crosses from above to below $w=-1$. 
This demonstrates a healthy phantom crossing in a purely 
shift symmetric Horndeski gravity model.

%%%%%%%%%%%%%%%%%%%%%%%% 

This is an important proof of principle. Thus one mechanism 
for phantom crossing in a healthy shift symmetric theory is 
to have $1/2<n\le1$. 
Of course the drawback is that there is no early matter 
(or radiation) dominated era so it is not {\it our\/} 
universe. However, one could imagine starting in the early 
universe with $0<n<1/2$, and having a somewhat more complicated 
theory than simple monomials that allows $n$ to evolve from 
$0<n_{\rm early}<1/2$ -- where the theory is healthy and matter 
domination holds -- to a later epoch of $1/2<n\le1$ where the 
theory has the right properties to cross the phantom divide. 

Evolution of $n$ was in fact studied in \cite{2512.03139,Calderon:2026hbr} 
and had the potential to cross the phantom divide. In 
\cite{Calderon:2026hbr} it was studied both in terms of $n(\ode)$ and 
the more fundamental $n(X)$. Adopting the $n(X)$ model of 
\cite{Calderon:2026hbr} we find that a purely shift symmetric theory 
with $0<n(X)<1/2$ at all times can be healthy and give one 
or even two phantom crossings -- the 
second mechanism for crossing is thus appropriate variation $n(X)$, and can preserve early matter domination. 

\begin{figure}
    \centering

\includegraphics[width=\linewidth]{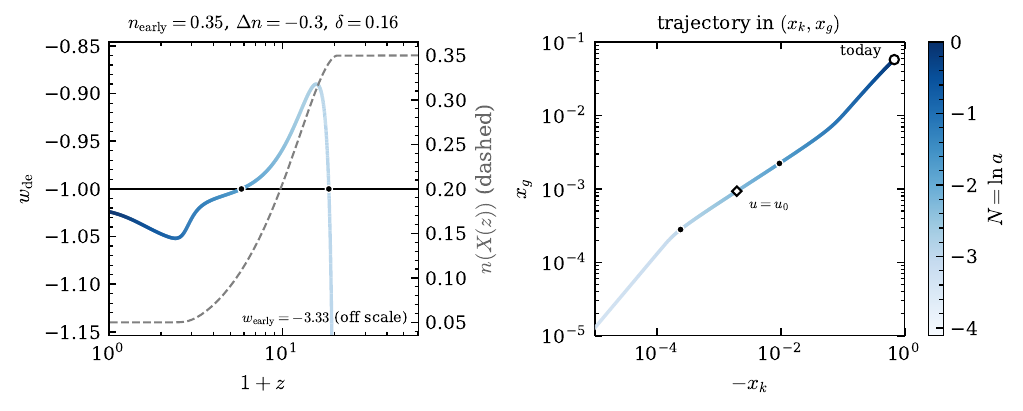}     
    \caption{
    A healthy phantom crossing for a shift symmetric model with $n(X)$ evolving from $\nearly=0.35$ at early times to $n_{\rm late}=0.05$ at late times. 
[Left panel] The dark energy equation of state successfully crosses $w=-1$. [Right panel] Phase space plot showing the trajectory of the evolving $n(X)$ model, and the $w=-1$ curves for the early time constant $n=1/4$ and late time constant $n=1/8$ cases. 
}
      \label{fig:nxcross}
\end{figure}

The sigmoid $n(X)$ of \cite{Calderon:2026hbr}, with $n(X)\equiv d\ln K/d\ln X$, 
is equivalent to replacing the $K\sim X^n$ 
monomial with almost as simple a form:  
\be 
K(X)\sim X^{\nearly}\,\left[1+\left(\frac{X}{X_c}\right)^{1/\dl}\right]^{\Delta n\,\dl}\ , 
\ee 
where $\Delta n=\nlate-\nearly$ gives the transition size, 
$X_c$ the transition timing, and $\dl$ the transition 
width. 
Note that the early time charge balance implies $m=(2\nearly-1)/4$, not 
$m=[2n(X)-1]/4$. 
Figure~\ref{fig:nxcross} demonstrates that phantom 
crossing indeed occurs (twice), and the model is healthy 
throughout its history. For illustration we use the 
values $n_{\rm early}=0.35$ and $n_{\rm late}=0.05$, 
i.e.\ $\Delta n=-0.3$. 
Note that using the early time behavior $X\sim a^{-6\we}$ 
we have $\ln(X_c/X_i)\approx-6\we\,\ln(a_c/a_i)\approx 30-40$; the figure uses $\ln(X_c/X_i)=34$. For the width $\dl=\Delta\ln X=6x_\phi\,\Delta\ln a$ 
we note that the phantom crossing occurs as $x_\phi$ 
approaches zero, so for a transition taking an e-fold of 
expansion we expect $\dl\approx0.1-0.2$; the figure uses 
$\dl=0.16$.

%%%%%%%%%%%%%%%%%%%%%%%% 

The phantom crossing can be viewed as a consequence of the 
rotation of the $w=-1$ parabola as $n$ changes, as seen in 
Figures~\ref{fig:flow-paperI} and Figures~\ref{fig:flown125375}. 
Basically the variation in $n$ sweeps the trajectory across the changing 
position of $w=-1$ in such a way as to enable crossing. 
Another way to view it is through Eq.~(\ref{eq:Qxgxk}). Since 
for $0<n<1/2$ we have $Q<0$, and $x_k<0$, then reducing $n$ 
diminishes the negative contribution from the $8n^2x_k$ term, 
which effectively raises $Q$ toward zero.

%%%%%%%%%%%%%%%%%%%%%%% 
\section{Conclusions} \label{sec:concl} 

A remarkably simple shift symmetry Horndeski gravity class 
with monomial kinetic and braiding terms possesses a rich 
phase space structure. By tracing this structure through 
dynamical systems analysis we identify the fixed points, 
attractors, and healthy regions free from ghosts and instability. 
In particular, dark energy can come to dominate the total 
energy density and give a de Sitter future state. The shift 
symmetry protects against troublesome high energy quantum 
loop corrections that bedevil almost all field dependent 
(non-shift symmetric) theories. 

The curve $w=-1$ is a parabola in the phase space and divides 
it into phantom and standard regions. Solving for the properties 
of the parabola and trajectories that intersect it, we determine 
analytically the conditions for crossing this phantom divide. 
Contrary to conventional thought, shift symmetric Horndeski 
gravity can cross $w=-1$ in a healthy manner. 
Note that many of the arguments against phantom crossing 
by shift symmetric modified gravity assumed zero charge, justifying it as the state of the 
de Sitter attractor (recall the charge density $C\propto a^{-3}$ so in the 
asymptotic future $C\to0$). However, $C=0$ implies 
$x_g=-2nx_k$ or $Q=x_g(x_g-2n)$ in our notation. This 
is simply a straight line in phase space between the 
fixed points $M$ and $dS$. It cannot cross the parabola 
of $Q=0$, just meet it at $dS$, $x_g=2n$. 

We identify two mechanisms for phantom crossing 
within a healthy shift symmetry theory with no potential or 
cosmological constant, and confirm their validity 
through full numerical solutions. 
One is replacing the $K\sim X^n$ 
monomial with nearly as simple a form having $n(X)$, 
and the second is taking initial conditions 
near the braiding point with $1/2<n\le1$. The latter 
does not give an early matter dominated era, but  
this might be dealt with by an appropriate $n(X)$ model. 
The first, with $0<n<1/2$, does has a standard early matter dominated era, 
and dark energy evolves from the phantom to $w>-1$ (as 
observations appear to prefer), and can even cross a second 
time to reach $w<-1$ again. 
Both mechanisms should be regarded here as proofs of principle for 
phantom crossing, not as models proposed to fit the current 
observational data. 

Finally, another approach to phantom crossing is to add 
field dependence in such a way as to preserve shift symmetry 
in the early universe, and hence protection from quantum 
corrections where it is needed most. This will be explored in  
future work \cite{Calderon:2026prep}. We emphasize that a major advantage 
of starting from the action, i.e.\ specifying the Horndeski 
terms, rather than parametrizing the time dependence of 
property functions, $\al_i(a)$, or gravitational coupling 
strengths, $\mu(a)$ and $\Sigma(a)$, is the dynamical autonomous 
system approach and its insights into trajectories, fixed 
points, and crossing $w=-1$, plus the possibility of computing 
the perturbative sector evolution as well.

%%%%%%%%%%%%%%%%%%%%%%%%%%%%% 
\section*{Acknowledgments}

We thank the Galileo Galilei Institute for hospitality during part of this work. 
R.C.\ is funded by the Czech Ministry of Education, Youth and Sports (MEYS) and European Structural and Investment Funds (ESIF) under project number CZ.02.01.01/00/22\_008/0004632. 
E.L.\ thanks the Simons Foundation for support as a Simons Visiting Scientist at GGI.

\bibliographystyle{JHEP}
\bibliography{refs}

\end{document}